\documentstyle[prl,twocolumn,aps,graphics,epsfig,floats]{revtex}

\newcommand{\be}{\begin{equation}}
\newcommand{\ee}{\end{equation}}
\newcommand{\bea}{\begin{eqnarray}}
\newcommand{\eea}{\end{eqnarray}}
\newcommand{\xe}{\mbox{$^{129}$}Xe}

\newcommand{\th}{\mbox{$\theta$}}
\newcommand{\oo}{\mbox{$^\circ$}}

\begin{document}
\draft

\title{Measurement of persistence in 1-D diffusion}
\author{Glenn P. Wong, Ross W. Mair, and Ronald L. Walsworth}
\address{Harvard-Smithsonian Center for Astrophysics, 60 Garden St., Cambridge, MA  02138}
\author{David G. Cory}
\address{Department of Nuclear Engineering, Massachusetts Institute of Technology, Cambridge,
MA  02139}

\date{\today}
\maketitle
\begin{abstract}
Using a novel NMR scheme we 
observed persistence in 1-D gas diffusion. 
Analytical approximations and numerical simulations have indicated that for an initially
random array of spins undergoing diffusion, the probability $p(t)$ that 
the average spin magnetization in a given region has
not changed sign (i.e., ``persists'') up to time $t$ follows a power law $t^{-\theta}$, where \th\
depends on the dimensionality of the system. 
Using laser-polarized $^{129}$Xe gas, we prepared an initial ``quasirandom''
1D array of spin magnetization and then monitored the ensemble's evolution
due to diffusion using real-time NMR imaging.
 Our measurements are consistent with
analytical and numerical predictions of \th\
$\approx$ 0.12.  

\end{abstract}

The dynamics of non-equilibrium systems is a field of great current
interest, including such topics as phase ordering in binary alloys,
uniaxial ferromagnets, and nematic liquid crystals, as well as coarsening of soap
froth and diffusion of inhomogeneous fluids
(\emph{e.g.} \cite{theoryreview:Bray}).
The evolving spatio-temporal
structures in these non-equilibrium systems depend crucially on the history
of the system's evolution and are not completely characterized by simple
measures such as two-time correlation functions.  Therefore, an important problem
in the study of non-equilibrium dynamics is the development of simple and
easily measurable quantities that give nontrivial information about the
history of the system's evolution.  The recently identified phenomenon of
``persistence'' may be such a quantity: it characterizes the statistics of first
passage events in spatially extended non-equilibrium systems
\cite{nontriv:Derr,nontriv:Bray,firstpass:Derr,2DPotts:Derr,Ising:Maj,diffpers:Derr,diffpers:Maj,Globalpers:Maj,persexp:Krug,perspoison:Lee,perspartial:Maj,diffpers:New,pers:Newman,pers:Jain,analyres:Sire,persexp:Kendon,persrand:Bray,breathfig:Marcos,exptpers:Tam,exptpers:Yurke}.
Persistence is being actively studied in statistical
physics; e.g., in the search for universal behavior in
non-equilibrium critical dynamics 
\cite{Globalpers:Maj,perspoison:Lee,persrand:Bray}.
Practically, persistence may be important in
determining what fraction of a system has reached a threshold condition as a function of time; e.g.,
in certain chemical reactions or disinfectant procedures.

Consider a non-equilibrium scalar field $\phi({\mathbf{x}},t)$ fluctuating in space and
time according to some dynamics (e.g., a random array of interdiffusing spins).  Persistence is
the probability $p(t)$ that at a fixed point in space the quantity
$[\phi({\mathbf{x}},t) - \langle \phi({\mathbf{x}},t) \rangle]$ has not changed sign up to
time $t$.
It has been found that this probability decays as a power law $p(t) \sim
t^{-\theta}$, where the persistence exponent $\theta$ is generally nontrivial and has been suggested as a new
universal dynamic critical exponent \cite{Globalpers:Maj,perspoison:Lee}.
This exponent depends both on the system dimensionality and the prevalent dynamics 
\cite{diffpers:Derr,diffpers:Maj}, and is
difficult to determine analytically due to the non-Markovian nature of the phenomena.  Although
$\theta$ has been calculated -- largely using numerical techniques -- for such systems as simple
diffusion \cite{diffpers:Derr,diffpers:Maj,diffpers:New}, the Ising model
\cite{firstpass:Derr,Ising:Maj,pers:Jain}, and the more generalized $q$-state Potts model
\cite{nontriv:Derr,firstpass:Derr,2DPotts:Derr}, few measurements of persistence have been
performed (see Table
\ref{tbl:thetavals}).  In particular,
 ``breath figures'' \cite{breathfig:Marcos}, 2-D soap froth \cite{exptpers:Tam},
and twisted nematic liquid crystals \cite{exptpers:Yurke} are the only systems for
which experimental results have been reported.  
Further experimental investigation is needed to probe the utility of persistence in studies of 
fundamental and applied non-equilibrium dynamics.

\begin{table}[b]
\caption{A sample of reported persistence exponents.  All values except those indicated
are derived from numerical simulations; ($^*$) denotes exact analytical results, ($^\dag$)
 experimental measurements, and ($^{\ddag}$) the result reported here.}  
\label{tbl:thetavals}
\begin{tabular}{cccc}
\textbf{Dim.} & \textbf{Diffusion} & \textbf{Ising} & $\mathbf{q}$\textbf{-Potts} \\
 \hline
1 & 0.12, \textbf{0.118}$^\ddag$ & 3/8$^*$, 0.35 &
$-\frac{1}{8}+\frac{2}{\pi^2}\left[\cos^{-1}\left(\frac{(2-q)}{\sqrt{2}q}\right)\right]^2
$$^*$\\
2 & 0.19 & 0.22, 0.19$^\dag$ & 0.86, 0.88$^\dag$ (large q) \\
3 & 0.24 & 0.26 & \\
\hline
\emph{refs} & 
\mbox{\emph{\cite{diffpers:Derr,diffpers:Maj,diffpers:New}}} &
\mbox{\emph{\cite{firstpass:Derr,Ising:Maj} \cite{exptpers:Yurke}$^\dag$}} &
\mbox{\emph{\cite{nontriv:Derr,firstpass:Derr,2DPotts:Derr} \cite{exptpers:Tam}$^\dag$} }
\end{tabular}
\end{table}

In this paper we present the first measurement of persistence in a system
undergoing diffusion (i.e., dynamics governed by Fick's Law $\dot{\phi} = D \phi''$).  
Our experiment is also the first to observe
persistence in one dimension (1-D).  We
employed a novel NMR technique to create an initial ``quasi-random'' spatial variation in the
nuclear spin magnetization of a sample of laser-polarized
$^{129}$Xe gas.  Subsequent 1-D NMR imaging allowed us to
monitor the temporal evolution of the ensemble.  We observed persistence by noting
mean magnetization sign changes at fixed locations of constant size (i.e., imaging pixels) as a function of
time.    Using a simple theory (the ``independent interval approximation'') and numerical
simulations, both Derrida
\emph{et al.} \cite{diffpers:Derr} and Majumdar \emph{et al.}
\cite{diffpers:Maj} independently found that 
\th\ $\approx 0.121$ for 1-D diffusion.  Newman and Toroczkai \cite{diffpers:New} found $\theta
\approx$ 0.125 in 1-D using an analytic expression for the diffusion persistence exponent.  Our
measurements are consistent with these calculations.

Recently, laser-polarized noble gas NMR has found wide application in both the physical and
biomedical sciences.  Examples include fundamental symmetry tests 
\cite{Bear_LLI}, 
probing the structure of porous media 
\cite{PorousPRL},
and imaging of the lung gas space
\cite{AlbertNature}.
These varied investigations, as well as the experiment reported here, exploit special
features of laser-polarized noble gas: the large nuclear spin polarization ($\sim$ 10\%) that can be
achieved with optical pumping techniques; the long-lived nuclear spin polarization of the spin-1/2 noble
gases $^{129}$Xe and $^3$He; and rapid gas-phase diffusion.


We performed laser-polarization of xenon gas using spin-exchange
optical pumping
\cite{RMP:Walker}.  
We filled a coated cylindrical glass cell \cite{OTS}
($\sim$ 9~cm long, 2~cm I.D.) with approximately 3 bar of xenon gas 
isotopically enriched to 90\% \xe, 0.5 bar of N$_2$ gas, and a small amount of 
Rb metal. We heated the sealed cell to
$\sim 100 ^\circ$C to create a significant Rb vapor.  
Optical pumping on the Rb D1 line was achieved with 15~W of
circularly-polarized 795~nm light (FWHM $\sim$ 3~nm) from a fiber-coupled laser diode array.
  After 20 minutes the \xe\ gas was routinely nuclear
spin-polarized to  1\% by spin-exchange collisions with the Rb vapor.  We next cooled
the cell to room temperature in a water bath and
placed the cell inside a homemade RF solenoid coil (2.5~cm diameter, 15~cm long, $Q
\sim$ 900) centered in a 4.7~T horizontal bore magnet (GE Omega/CSI spectrometer/imager) with
\xe\ Larmor frequency = 55.345~MHz.  To allow the gas temperature to reach equilibrium, we left
the cell in place for 20~minutes before starting the persistence measurements.  
At equilibrium under these conditions, the \xe\ polarization decay time constant
($T_1$) was in excess of 3 hours, with a \xe\ diffusion coefficient of 0.0198~cm$^2$/s 
\cite{TimeGasDiff}. (Note that changes in the sample gas pressure, and hence the diffusion coefficient, simply
cause a rescaling of the time variable and do not affect the persistence power law $p(t) \sim t^{-\theta}$ 
\cite{diffpers:Maj}.) 

\begin{figure}[t]
\begin{center}
\epsfig{file=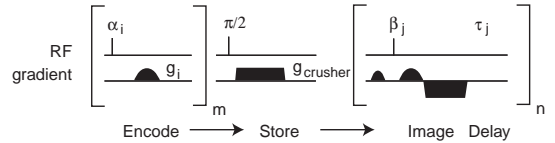, width=2.8in}
\caption{NMR pulse sequence used to encode a 1-D ``quasi-random''
pattern on the average magnetization of laser-polarized \xe\ gas. Temporal evolution of the
magnetization pattern is monitored with $n$ repetitions of a 1-D FLASH imaging routine \protect\cite{FLASH}. 
For example, with $m = 8$ encoding RF pulse/gradient pairs, the encoding pulse angles
$\alpha_i$ = [30\oo, 35\oo, 37\oo, 41\oo, 45\oo, 50\oo, 63.5\oo, and 90\oo] while the encoding gradient
amplitudes $g_i$ were chosen randomly.  The imaging pulse angle $\beta_j$ was fixed at 8\oo\ and
the diffusion times $\tau_j$ were varied from 2.4 ms up to $\sim$ 2 s.  The encoding gradients
and the transverse magnetization dephasing
``crusher'' gradient were pulsed for 1 and 20 ms, respectively.  Imaging gradients were applied 
for a total of 7.56 ms.  The maximum gradient available was 6.7 G/cm.}
\label{fig:pulseprog}
\end{center}
\end{figure}

The NMR pulse sequence we used to observe persistence in laser-polarized \xe\ gas
diffusion is shown schematically in Fig.~\ref{fig:pulseprog}.  The first portion of the pulse
sequence encodes a 1-D ``quasi-random'' pattern on the transverse magnetization of the
laser-polarized \xe\ gas sample by using $m$ pairs of variable-strength RF and magnetic-field-gradient pulses,
repeated in rapid succession (see Fig.~\ref{fig:pulseprog}).  Each pair of RF and gradient pulses adds different
spatial Fourier components to the 1-D transverse magnetization pattern, with wavenumbers given by the gradient pulse
area and Fourier component amplitudes set by the RF pulse area. Next, a $\pi/2$
RF pulse ``stores'' this quasi-random 1-D magnetization distribution along the longitudinal ($z$)
direction while a subsequent strong (crusher) gradient pulse dephases any remaining
transverse magnetization.   The quasi-random longitudinal magnetization distribution then evolves
with time due to diffusion and is monitored by a series of
1-D FLASH (Fast Low Angle SHot) NMR images  \cite{FLASH,persistence_pulseseq} (see Fig.~\ref{fig:pulseprog}).
 
The initial pattern of longitudinal \xe\ magnetization is quasi-random
in that there must be a minimum length scale to the induced variations in the \xe\ magnetization, i.e., a maximum
wavenumber in the pattern, for there to be sufficient NMR signal for useful imaging.  (This minimum
length scale was $\approx 0.6$~ mm in our experiment.)  Nevertheless, at longer length scales the induced pattern
must be random enough that persistence behavior can be expected.  Ideally, $\langle \phi(x,0) \phi(x',0) \rangle =
\delta(x-x')$; however, calculations indicate that it is sufficient for the initial condition correlator to decrease
faster than $| x-x' |^{-1}$ for 1-D persistence \cite{diffpers:Maj}.
We found that six to eight RF/gradient pulse pairs ($m =$ 6--8) were optimal for the desired
quasi-random 1-D patterning of the \xe\ magnetization.
$m < 6$ resulted in a pattern that was not sufficiently
random, while $m > 8$ significantly reduced the signal-to-noise ratio
(SNR) of the NMR images.  
The requirement of $m \geq 6$ is supported by
numerical calculations in which we modeled the NMR encoding sequence and simulated the subsequent
gas diffusion using a finite difference first-order forward Euler scheme
\cite{diffpers:Derr,numrecipe}: we found persistence behavior (i.e., $p(t) \sim t^{-\theta}$)
only when $m \geq 6$.  The requirement of $m \leq 8$ was set by the available NMR signal (i.e., the finite \xe\
magnetization), the necessity of rapid data acquisition to avoid excessive diffusion during the imaging
sequence itself, the limitation of approximately 2$\pi \times$(0.6 mm)$^{-1}$ for the maximum wavenumber, and the
maximum imaging gradient strength available.

\begin{figure}
\begin{center}
\epsfig{file=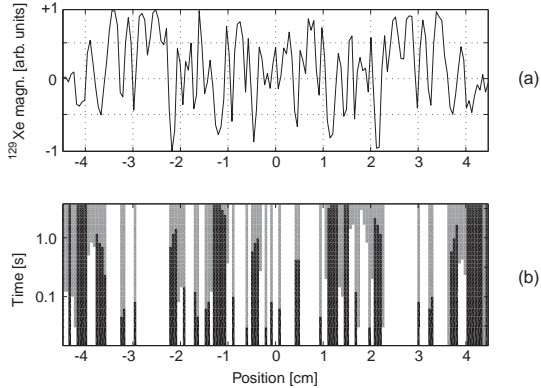, width=2.8in}
\caption{(a) Typical quasi-random initial pattern of \xe\ magnetization resulting from 8 encoding RF
pulse/gradient pairs.  (b) 32 images of the evolving magnetization pattern were acquired at logarithmically
increasing times. White (black) indicates average 
positive (negative) magnetizations for each of the 150 fixed-location imaging pixels of width $\approx$ 0.6~mm. 
Gray indicates pixels whose magnetization has changed sign at least once.  The persistence exponent is determined
from the growing fraction of gray pixels as a function of time.}
\label{fig:tempevol}
\end{center}
\end{figure}

For NMR imaging, we used a field of view (FOV) of
31.5 cm with 0.6 mm resolution, which thus divided the 9 cm cell into about 150 imaging pixels, each
corresponding to a discernible spatial region of fixed size and location.  We typically employed 
8\oo\ excitation RF pulse angles and acquired 32 1-D images spaced logarithmically in time from
$\sim$ 3 ms to 5 s for a single experimental run.  Longitudinal magnetization depletion due to imaging
was highly uniform across the sample and did not affect the persistence measurement, since the relative
magnetization amplitudes in neighboring imaging pixels was unchanged.
  An example of the
images acquired in a typical run are shown in Fig.~\ref{fig:tempevol}.  
For each pixel, we derived average magnetizations (aligned or anti-aligned to the main magnetic field) from the
phase information contained in the time-domain NMR image data, and spatial positions from the frequency information 
\cite{NMR:phaseit}. An experimental run thus provided a record of the \xe\ gas magnetization in each pixel as a
function of time proceeding from the initial quasi-random pattern to the equilibrium
condition of homogeneous (near-zero) polarization.

\begin{figure}
\begin{center}
\epsfig{file=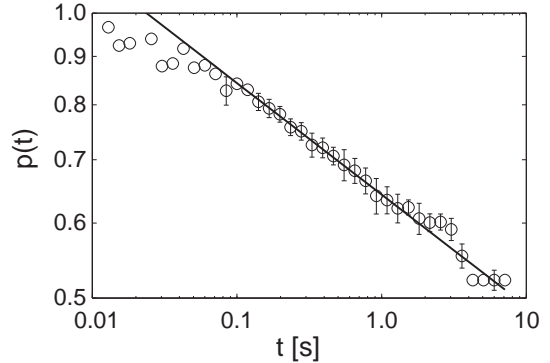, width=2.8in}
\caption{A log-log plot of $p(t)$, the fraction of regions (pixels) with $^{129}$Xe spin magnetization 
that had not changed sign up to a time $t$, representing the sum of
$\sim$ 30 different experimental runs.  The solid line is a weighted linear least-squares fit to the data
for $0.1$~s $\leq t \leq 1$~s, and yields $\theta = 0.118 \pm 0.008$.  Error bars are derived
from the number of pixels with amplitudes close to the image noise level and are
shown when they exceed the plot symbol diameter.}
\label{fig:pers}
\end{center}
\end{figure}

To measure persistence, we noted the sign of the \xe\ magnetization in each fixed
spatial region (i.e., in each 1-D image pixel) and counted how many remained unchanged as a
function of time.  We equated the probability $p(t)$
with the fraction of pixels that had
not changed sign up to time $t$.  We
chose $t=0$ to coincide with the first image   
and assigned the time index for each image to be the start time of the imaging RF
pulse.  Images with SNR $< 40$ were excluded from the data to minimize uncertainty in pixel sign
changes.  We conducted about 30 experiments with image SNR $> 40$, each with a unique set of
randomly chosen encoding magnetic field gradients \{$g_i$\}. 
We observed that pixel sign changes occurred smoothly, so it was unlikely we missed sign changes with an error of
more than one step in the imaging sequence.
 We employed two averaging schemes to combine
the results from different experimental runs. In the first method, we used a linear least-squares fit
of $\log[p(t)]$ vs.\ $\log[t]$ for each run, resulting in a distribution of power law exponents
with a weighted mean \th\ $= 0.119 \pm 0.048$.  With our numerical simulations of quasi-random initial
conditions, we found that this averaging scheme results in a gaussian distribution of exponents with a mean
value $\theta
\approx$ 0.12 in agreement with previous calculations for 1-D diffusion 
\cite{diffpers:Derr,diffpers:Maj,diffpers:New} and our experimental results. 
In the second averaging scheme, we 
combined the data from all experimental runs; hence $p(t)$ represented the fraction of total
pixels from all experiments that had not changed sign up to time $t$.
We found $p(t) \sim t^{-\theta}$ with \th\ $= 0.118 \pm 0.008$ for 0.1~s $\leq t \leq$\ 1~s.
Figure \ref{fig:pers} shows a log-log plot of $p(t)$ vs.\ $t$ when the data is averaged using
this method.

\begin{figure}
\begin{center}
\epsfig{file=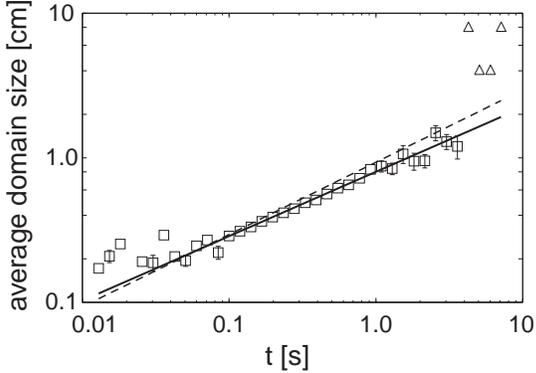, width=2.8in}
\caption{The average size $L$ of domains of adjacent pixels with uniform magnetization orientation,
as a function of time $t$, derived from all experimental runs.  For
$0.1$~s $ \leq t \leq 1$~s, $L \sim t^\alpha$ where $\alpha$ = $0.45 \pm 0.02$ (solid line). 
The dotted line shows the expected $L \sim t^{1/2}$ 
behavior for an infinite system.  The error in $L$ is shown where it exceeds
the plot symbol size.  The finite size limit on $L$ is evident in the four late-time points
($\triangle$), which were taken from the only two runs with sufficient SNR at long times.}
\label{fig:doms}
\end{center}
\end{figure}

The observed deviations from power law behavior for $t \lesssim 0.1$~s and $t
\gtrsim 1$~s are explained by image resolution and finite size effects,
respectively.  At short times persistence is not observed because \xe\ atoms have not yet diffused on average across
a single 1-D image pixel ($\approx 0.6$~mm).  The relevant diffusion time is $(0.6$~mm$)^2/(2 D_{Xe}) \approx 0.1$~s
for our typical experimental conditions.  At long times, the coarsening of the \xe\ magnetization pattern by
diffusion results in large ``domains'' of adjacent pixels with the same 
sign of the magnetization.  Our simulations indicate that persistence is not observed when there are few domains in
the finite size sample because the number of magnetization boundaries is greatly reduced;
hence the rate of pixel sign-changing (i.e., the growth of
the gray area in Fig.~\ref{fig:tempevol}) is curtailed.
Both the short and long-time deviations are seen in Fig.~\ref{fig:doms}, where the average
length $L$ of like-signed magnetization domains from all experimental 
runs is plotted against time.  For $0.1$~s
$ \lesssim t \lesssim 1$~s, our data are in reasonable agreement with the expected power law $L \sim
t^{1/2}$ for diffusion \cite{theoryreview:Bray}.   For $t \gtrsim 1$~s, we find $L \gtrsim 1$~cm, implying there are
typically less than 10 magnetization boundaries in the 9 cm long sample cell.


In conclusion, we experimentally measured a persistence exponent \th\ $\approx$ 0.12 for 1-D
diffusion, consistent with analytical and numerical studies.  We performed the measurement
using a novel NMR scheme with laser-polarized \xe\ gas which allowed us to both encode a
spatially ``quasi-random'' magnetization pattern and monitor its evolution 
over several seconds.  We also observed the effect of finite sample size for long-time diffusion.
In future work the experimental technique employed in this study may allow measurements of
persistence in 2 and 3-D diffusion, in heterogeneous systems (e.g., porous media) infused
with noble gas, and in `patterns' \cite{persnoneq:Maj}.

The authors thank Satya Majumdar, Michael
Crescimanno, and Lukasz Zielinski for useful discussions.
This work was supported by NSF Grant No.~CTS-9980194, NASA Grant No.~NAG9-1166, 
and the Smithsonian Institution
Scholarly Studies Program.

%


\begin{thebibliography}{10}

\bibitem{theoryreview:Bray}
A.~J. Bray, Adv.\ Phys. {\bf 32},  357  (1994).

\bibitem{nontriv:Derr}
B. Derrida, A. Bray, and C. Godr\`{e}che, J.\ Phys.\ A. {\bf 27},  L357  (1994).

\bibitem{nontriv:Bray}
A. Bray, B. Derrida, and C. Godr\`{e}che, Eur.\ Phys.\ Lett. {\bf 27},  175
  (1994).

\bibitem{firstpass:Derr}
B. Derrida, V. Hakim, and V. Pasquier, Phys.\ Rev.\ Lett. {\bf 75},  751
  (1995).

\bibitem{2DPotts:Derr}
B. Derrida, P.~M.~C. de~Oliveira, and D. Stauffer, Physica A {\bf 224},  604
  (1996).

\bibitem{Ising:Maj}
S.~N. Majumdar and C. Sire, Phys.\ Rev.\ Lett. {\bf 77},  1420  (1996).

\bibitem{diffpers:Derr}
B. Derrida, V. Hakim, and R. Zeitak, Phys.\ Rev.\ Lett. {\bf 77},  2871
  (1996).

\bibitem{diffpers:Maj}
S.~N. Majumdar, C. Sire, A.~J. Bray, and S.~J. Cornell, Phys.\ Rev.\ Lett. {\bf
  77},  2867  (1996).

\bibitem{Globalpers:Maj}
S.~N. Majumdar, A. Bray, S. Cornell, and C. Sire, Phys.\ Rev.\ Lett. {\bf 77},
  3704  (1996).

\bibitem{persexp:Krug}
J. Krug, H. Kallabis, S.~N. Majumdar, S.~J. Cornell, A.~J. Bray, and C. Sire,
  Phys.\ Rev.\ E {\bf 56},  2702  (1997).

\bibitem{perspoison:Lee}
B.~P. Lee and A.~D. Rutenberg, Phys.\ Rev.\ Lett. {\bf 79},  4842  (1997).

\bibitem{perspartial:Maj}
S.~N. Majumdar and A.~J. Bray, Phys.\ Rev.\ Lett. {\bf 81},  2626  (1998).

\bibitem{diffpers:New}
T.~J. Newman and Z. Toroczkai, Phys.\ Rev.\ E {\bf 58},  R2685  (1998).

\bibitem{pers:Newman}
C.~M. Newman and D.~L. Stein, Phys.\ Rev.\ Lett. {\bf 82},  3944  (1999).

\bibitem{pers:Jain}
S. Jain, Phys.\ Rev.\ E {\bf 60},  R2445  (1999).

\bibitem{analyres:Sire}
C. Sire, S.~N. Majumdar, and A. R{\"u}dinger, Phys.\ Rev.\ E {\bf 60},  1258
  (2000).

\bibitem{persexp:Kendon}
V.~M. Kendon, M.~E. Cates, and J.-C. Desplat, Phys.\ Rev.\ E {\bf 61},  4029
  (2000).

\bibitem{persrand:Bray}
A.~J. Bray, Phys.\ Rev.\ E {\bf 62},  103  (2000).

\bibitem{breathfig:Marcos}
M. Marcos-Martin, D. Beysens, J.~P. Bouchaud, C. Godr\`{e}che, and I.
  Yekutieli, Physica A {\bf 214},  396  (1995).

\bibitem{exptpers:Tam}
W.~Y. Tam, R. Zeitak, K.~Y. Szeto, and J. Stavans, Phys.\ Rev.\ Lett. {\bf 78},
   1588  (1997).

\bibitem{exptpers:Yurke}
B. Yurke, A.~N. Pargellis, S.~N. Majumdar, and C. Sire, Phys.\ Rev.\ E {\bf
  56},  R40  (1997).

\bibitem{Bear_LLI}
D. Bear, R.~E. Stoner, R.~L. Walsworth, V.~A. Kostelecky, and C.~D. Lane,
  Phys.\ Rev.\ Lett. {\bf 85},  5038  (2000).

\bibitem{PorousPRL}
R.~W. Mair, G.~P. Wong, D. Hoffmann, M.~D. H{\"{u}}rlimann, S. Patz, L.~M.
  Schwartz, and R.~L. Walsworth, Phys.\ Rev.\ Lett. {\bf 83},  3324  (1999).

\bibitem{AlbertNature}
M.~S. Albert, G.~D. Cates, B. Driehuys, W. Happer, B. Saam, C.~S. Springer,
  Jr., and A. Wishnia, Nature {\bf 370},  199  (1994).

\bibitem{RMP:Walker}
T.~G. Walker and W. Happer, Rev.\ Mod.\ Phys. {\bf 69},  629  (1997).

\bibitem{OTS}
We used a wall coating of octadecyltrichlorosilane (OTS) to reduce Xe-wall
  interactions and hence increase longitundinal relaxation times.

\bibitem{TimeGasDiff}
R.~W. Mair, D.~G. Cory, S. Peled, C.-H. Tseng, S. Patz, and R.~L. Walsworth,
  J.\ Mag.\ Res. {\bf 135},  478  (1998).

\bibitem{FLASH}
A. Haase, J. Frahm, D. Matthaei, W. H{\"{a}}nicke, and K.-D. Merboldt, J.\
  Mag.\ Res. {\bf 67},  258  (1986).

\bibitem{persistence_pulseseq}
A more detailed description of the NMR pulse sequence used in this experiment
  will be presented elsewhere.

\bibitem{numrecipe}
W.~H. Press, B.~P. Flannery, S.~A. Teukolsky, and W.~T. Vetterling, {\em
  Numerical Recipes in {C}} (Cambridge University Press, Cambridge, U.K.,
  1988).

\bibitem{NMR:phaseit}
C.~B. Ahn and Z.~H. Cho, IEEE Trans.\ Med.\ Imag. {\bf MI-6},  32  (1987).

\bibitem{persnoneq:Maj}
S.~N. Majumdar, Curr.\ Sci. {\bf 77},  370  (1999).

\end{thebibliography}

\end{document}